\documentclass{article}

\usepackage[preprint]{neurips_2024}
\usepackage[utf8]{inputenc}
\usepackage[T1]{fontenc}
\usepackage{hyperref}
\usepackage{url}
\usepackage{booktabs}
\usepackage{amsfonts}
\usepackage{amsmath}
\usepackage{amssymb}
\usepackage{nicefrac}
\usepackage{microtype}
\usepackage{graphicx}
\usepackage{subcaption}
\usepackage{algorithm}
\usepackage{algorithmic}
\usepackage{natbib}

\title{Recovering Sparse Neural Connectivity from Partial Measurements:\\
A Covariance-Based Approach with Granger-Causality Refinement}

\author{
  Quilee Simeon \\
  Massachusetts Institute of Technology \\
  \texttt{qsimeon@mit.edu}
}

\begin{document}

\maketitle

\begin{abstract}
Inferring the connectivity of neural circuits from incomplete observations is a fundamental challenge in neuroscience.
We present a covariance-based method for estimating the weight matrix of a recurrent neural network from sparse, partial measurements across multiple recording sessions.
By accumulating pairwise covariance estimates across sessions where different subsets of neurons are observed, we reconstruct the full connectivity matrix without requiring simultaneous recording of all neurons.
A Granger-causality refinement step enforces biological constraints via projected gradient descent.
Through systematic experiments on synthetic networks modeling small brain circuits, we characterize a fundamental \emph{control-estimation tradeoff}: stimulation aids identifiability but disrupts intrinsic dynamics, with the optimal level depending on measurement density.
We discover that the ``incorrect'' linear approximation acts as implicit regularization---outperforming the oracle estimator with known nonlinearity at all operating regimes---and provide an exact characterization via the Stein--Price identity.
\end{abstract}

\section{Introduction}
\label{sec:intro}

A central goal in systems neuroscience is to map the connectivity of neural circuits from functional measurements \citep{power2011functional}.
In small brain circuits---such as \textit{C.~elegans} \citep{white1986structure, cook2019whole}, larval zebrafish \citep{ahrens2013whole}, and cortical organoids \citep{quadrato2017cell}---the number of neurons is tractable ($10^2$--$10^4$) but simultaneous recording of the entire circuit remains difficult.
This creates a fundamental inverse problem: given partial, noisy measurements of neural dynamics across multiple sessions, can we recover the full connectivity matrix?

This problem embodies a \emph{control-estimation tradeoff} related to classical notions of persistent excitation in system identification \citep{ljung1998system}: stimulating neurons with random perturbations aids connectivity estimation by exciting network modes, but disrupts the intrinsic dynamics that make the circuit biologically relevant.
An ideal experimental protocol must balance identifiability against preservation of natural dynamics.
Furthermore, real circuits possess intrinsic oscillatory patterns---central pattern generators (CPGs) \citep{marder2001central}---whose detailed structure is unknown to the experimenter.
These represent \emph{unmodeled autonomous dynamics} that influence the observed activity.

We formalize this as a matrix recovery problem from partial observations.
Consider a network of $N$ neurons with connectivity matrix $W \in \mathbb{R}^{N \times N}$, governed by discrete-time recurrent dynamics \citep{sussillo2009generating}, where $W_{ij}$ represents the synaptic weight from neuron $j$ to neuron $i$.
In each recording session, we observe a subset of neurons and stimulate a (possibly different) subset.
Our key insight is that pairwise covariance statistics can be accumulated across sessions: whenever neurons $i$ and $j$ are simultaneously observed, their covariance contributes to estimating the $(i,j)$ block of $W$.

\paragraph{Contributions.}
Our method is general, applying to any network where the same connectivity is measured across multiple partial-observation sessions.
\begin{itemize}
    \item We derive a covariance-based estimator $\hat{W} = \Sigma_{x_{t+1},x_t} \Sigma_{x_t,x_t}^{-1}$ for connectivity recovery from partial measurements and provide identifiability conditions.
    \item We introduce a Granger-causality-inspired \citep{granger1969investigating} refinement via projected gradient descent that enforces biological constraints (sparsity, non-negativity, no autapses).
    \item We discover an \emph{implicit regularization} effect: the ``incorrect'' linear approximation yields a better-conditioned estimator than the oracle with known nonlinearity, analogous to James--Stein shrinkage \citep{james1961estimation}.
    \item We systematically characterize the tradeoff between stimulation strength and measurement density, revealing that optimal stimulation depends on the fraction of observed neurons.
\end{itemize}

\section{Problem Formulation}
\label{sec:formulation}

\subsection{Dynamical System Model}

We model the neural circuit as a discrete-time recurrent dynamical system:
\begin{equation}
    x_{t+1} = W \phi(x_t) + b_t
    \label{eq:dynamics}
\end{equation}
where $x_t \in \mathbb{R}^N$ is the neural state, $W \in \mathbb{R}^{N \times N}$ is the connectivity matrix, $\phi: \mathbb{R} \to \mathbb{R}$ is an element-wise nonlinearity (e.g., $\tanh$), and $b_t \in \mathbb{R}^N$ is the total input comprising both extrinsic stimulation and intrinsic central pattern generator (CPG) drive.

For stability, we require $\rho(W) \leq 1$, where $\rho(\cdot)$ denotes the spectral radius.

\subsection{Observation Model}

In session $k$, we observe a subset $\mathcal{M}_k \subseteq \{1, \ldots, N\}$ of neurons and stimulate a subset $\mathcal{S}_k$.
The observed activity is:
\begin{equation}
    y_t^{(k)} = P_{\mathcal{M}_k} x_t + \epsilon_t, \qquad \epsilon_t \sim \mathcal{N}(0, \sigma_\epsilon^2 I)
\end{equation}
where $P_{\mathcal{M}_k}$ is a projection onto the observed neurons and $\epsilon_t$ is i.i.d.\ measurement noise.
Our experiments test both noiseless ($\sigma_\epsilon = 0$) and noisy ($\sigma_\epsilon > 0$) regimes; the method is robust to moderate noise (Section~\ref{sec:discussion}).

\subsection{The Estimation Problem}

Given $K$ sessions with observations $\{y_t^{(k)}\}_{t=1}^{T_k}$ for $k = 1, \ldots, K$, recover $W$.

\section{Methods}
\label{sec:methods}

\subsection{Covariance-Based Estimator}

Under the linear approximation $\phi(x) \approx x$ (valid when $\|x\|$ is small), Eq.~\eqref{eq:dynamics} simplifies to $x_{t+1} \approx W x_t + b_t$.
Assuming statistical independence between inputs $b_t$ and states $x_t$ (an approximation we revisit critically in Section~\ref{sec:discussion}), the cross-covariance of the \emph{true} states satisfies:
\begin{equation}
    \Sigma_{x_{t+1}, x_t} = W \Sigma_{x_t, x_t} + \underbrace{\Sigma_{b_t, x_t}}_{\approx 0}
    \label{eq:covariance}
\end{equation}

In the noiseless case ($\sigma_\epsilon = 0$, so $y_t = x_t$), this directly yields the estimator $\hat{W} = \Sigma_{y_{t+1}, y_t} \Sigma_{y_t, y_t}^{-1}$.
When observation noise is present ($\sigma_\epsilon > 0$), the observed covariances differ from the true state covariances.
Since $\epsilon_t$ is i.i.d.\ and independent of $x_t$:
\begin{align}
    \Sigma_{y_t, y_t} &= \Sigma_{x_t, x_t} + \sigma_\epsilon^2 I \label{eq:noisy_cov} \\
    \Sigma_{y_{t+1}, y_t} &= \Sigma_{x_{t+1}, x_t} \quad \text{(noise at $t{+}1$ independent of $y_t$)}
\end{align}
The estimator computed from observed data is therefore:
\begin{equation}
    \hat{W} = \Sigma_{y_{t+1}, y_t} \bigl(\Sigma_{y_t, y_t}\bigr)^{-1} = W \Sigma_{x,x} \bigl(\Sigma_{x,x} + \sigma_\epsilon^2 I\bigr)^{-1}
    \label{eq:estimator}
\end{equation}
This is a \emph{ridge-regularized} version of the true $W$: observation noise shrinks the estimate toward zero, with $\sigma_\epsilon^2$ playing the role of the ridge parameter.
For small noise ($\sigma_\epsilon \ll \sigma_{\min}(\Sigma_{x,x})$), the effect is negligible; for large noise, the estimate is over-regularized.

\paragraph{Accumulation across sessions.}
For each pair $(i,j)$, we compute empirical covariances from the \emph{observed} data $y_t$ only in sessions where both neurons $i$ and $j$ are simultaneously measured.
Let $S_{ij}^{(k)} = \mathbf{1}[i \in \mathcal{M}_k \wedge j \in \mathcal{M}_k]$ indicate co-measurement.
The accumulated covariance is:
\begin{equation}
    \hat{\Sigma}_{ij} = \frac{\sum_{k=1}^{K} S_{ij}^{(k)} \hat{\Sigma}_{ij}^{(k)}}{\sum_{k=1}^{K} S_{ij}^{(k)}}
\end{equation}
where $\hat{\Sigma}_{ij}^{(k)}$ is the empirical covariance of the observed $y_t$ from session $k$.

\paragraph{Error analysis.}
The estimation error decomposes into two terms:
\begin{equation}
    \hat{W} - W = \underbrace{W\bigl[\Sigma_{\phi(x),x} - \Sigma_{x,x}\bigr]\Sigma_{x,x}^{-1}}_{E_1:\text{ model mismatch}} + \underbrace{\Sigma_{b,x}\Sigma_{x,x}^{-1}}_{E_2:\text{ input correlation}}
    \label{eq:error}
\end{equation}
For $\phi = \tanh$ and Gaussian states $x \sim \mathcal{N}(0, \Sigma)$, the Stein--Price identity \citep{price1958useful, stein1981estimation} gives an \emph{exact} characterization.
Since $\mathbb{E}[\tanh(x_i) x_j] = \Sigma_{ij} \, \mathbb{E}[\mathrm{sech}^2(x_i)]$, we have:
\begin{equation}
    \Sigma_{\phi(x),x} = D' \Sigma_{x,x}, \qquad D' = \mathrm{diag}\!\big(\mathbb{E}[\mathrm{sech}^2(x_i)]\big)
    \label{eq:price}
\end{equation}
where $0 < D'_{ii} < 1$ for $\sigma_i > 0$.
Since $E_1 = W(D' - I)\Sigma_{x,x}\Sigma_{x,x}^{-1} = W(D' - I)$, the mismatch error simplifies to $\|E_1\|_F = \|W(D'-I)\|_F \leq \|W\|_F \max_i(1 - \mathbb{E}[\mathrm{sech}^2(x_i)])$.
For small state variance, $1 - \mathbb{E}[\mathrm{sech}^2(x_i)] \approx \sigma_i^2$, so the error is $O(\max_i \sigma_i^2)$---substantially tighter than generic Lipschitz bounds.
Crucially, $\|\Sigma_{x,x}^{-1}\|_2$ amplifies both error sources: ill-conditioned covariance (from sparse measurement) degrades recovery even with perfect model knowledge.

\subsection{Granger-Causality Refinement}

The raw estimator $\hat{W}$ may violate biological constraints.
We refine it using projected gradient descent that enforces:

\begin{enumerate}
    \item \textbf{No self-connections}: $W_{ii} = 0$ for all $i$ (no autapses).
    \item \textbf{Granger non-causality}: $W_{ij} = 0$ where $\Sigma_{x_t, x_t}(i,j) > \Sigma_{x_{t+1}, x_t}(i,j)$, indicating that the lagged covariance does not exceed the contemporaneous covariance.
    \item \textbf{Non-negativity}: $W_{ij} \geq 0$ (when applicable, e.g., for excitatory-only networks).
\end{enumerate}

The optimization solves:
\begin{equation}
    \min_A \|A - \hat{\Sigma}_{x_{t+1}, x_t}\|_F^2 \quad \text{s.t.} \quad A \Sigma_{x_t, x_t}^{-1} \in \mathcal{C}
    \label{eq:optimization}
\end{equation}
where $\mathcal{C}$ is the constraint set.
We alternate between gradient steps on $A$ and projection of $W = A \Sigma_{x_t, x_t}^{-1}$ onto $\mathcal{C}$.

\subsection{Full Pipeline}

\begin{figure}[h]
\centering
\includegraphics[width=\linewidth]{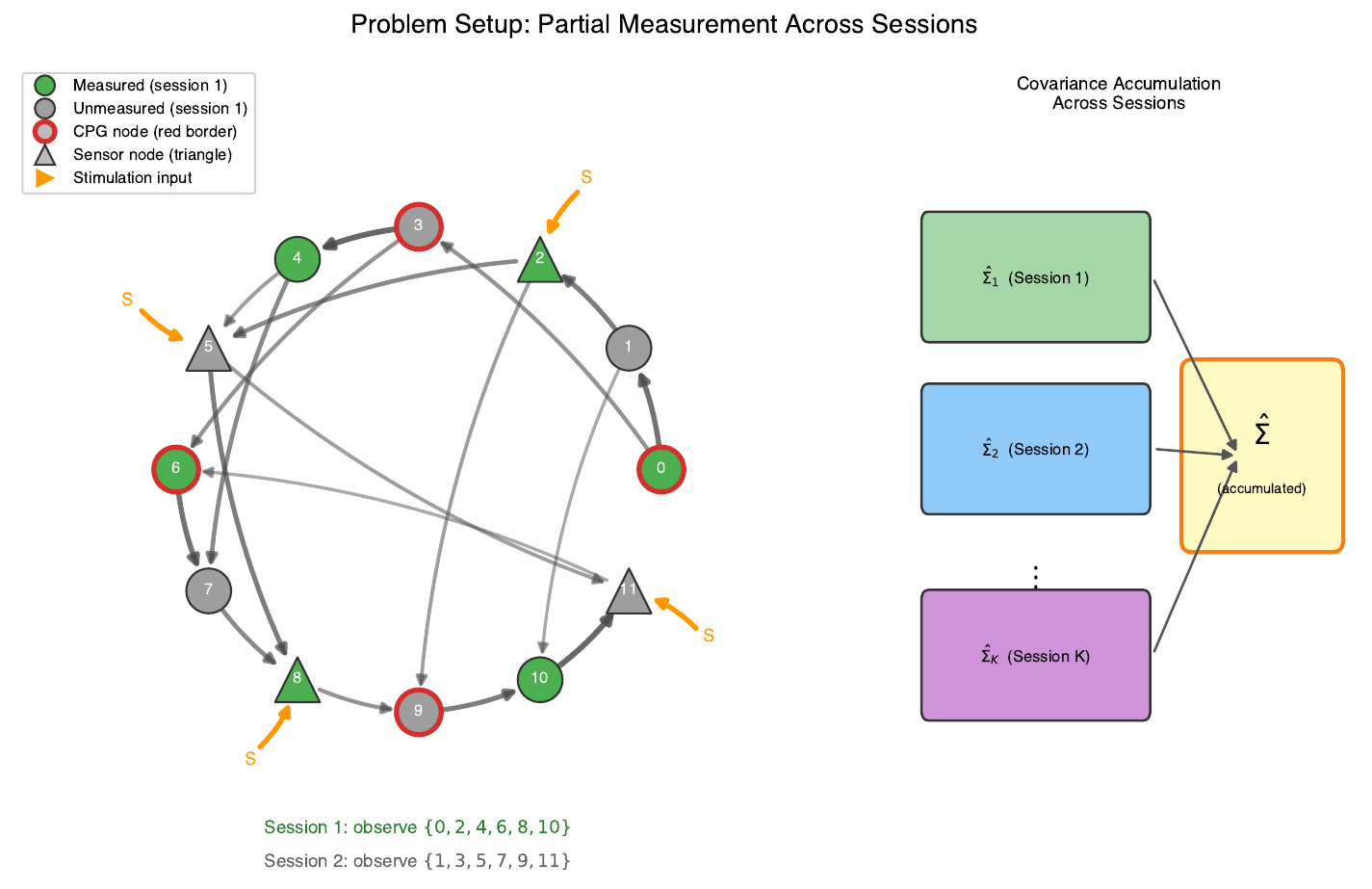}
\caption{\textbf{Problem setup: partial measurement across recording sessions.} \textbf{(Left)} A network of $N=12$ neurons with directed weighted connectivity $W$. In Session~1, a subset of neurons (green circles) is measured while the rest (gray) are unobserved. Nodes with red borders are central pattern generator (CPG) drivers; triangle-shaped nodes are sensor neurons receiving extrinsic stimulation (orange arrows labeled ``S''). \textbf{(Right)} Covariance accumulation across $K$ sessions: each session yields a partial covariance matrix $\hat{\Sigma}_k$ covering only co-observed neuron pairs. These are averaged element-wise to reconstruct the full accumulated covariance $\hat{\Sigma}$, from which $W$ is estimated via Eq.~\eqref{eq:estimator}.}
\label{fig:schematic}
\end{figure}

The complete method is summarized in Algorithm~\ref{alg:pipeline} and illustrated schematically in Figure~\ref{fig:schematic}.

\begin{algorithm}[h]
\caption{Covariance Accumulation + Granger Refinement}
\label{alg:pipeline}
\begin{algorithmic}[1]
\REQUIRE Sessions $\{(y_t^{(k)}, \mathcal{M}_k)\}_{k=1}^K$
\ENSURE Estimated connectivity $\hat{W}$
\STATE Initialize $\hat{\Sigma}_{y_{t+1},y_t} \leftarrow 0$, $\hat{\Sigma}_{y_t,y_t} \leftarrow 0$, $C \leftarrow 0$ \COMMENT{accumulators}
\FOR{$k = 1, \ldots, K$}
    \STATE Compute session covariances from observed data $\{y_t^{(k)}\}_{t=1}^{T_k}$
    \STATE $S \leftarrow \mathbf{1}_{\mathcal{M}_k} \mathbf{1}_{\mathcal{M}_k}^T$ \COMMENT{co-measurement mask}
    \STATE $\hat{\Sigma}_{y_{t+1},y_t} \mathrel{+}= S \odot \hat{\Sigma}^{(k)}_{y_{t+1},y_t}$; \quad $\hat{\Sigma}_{y_t,y_t} \mathrel{+}= S \odot \hat{\Sigma}^{(k)}_{y_t,y_t}$
    \STATE $C \mathrel{+}= S$
\ENDFOR
\STATE $\hat{\Sigma}_{y_{t+1},y_t} \leftarrow \hat{\Sigma}_{y_{t+1},y_t} \oslash C$; \quad $\hat{\Sigma}_{y_t,y_t} \leftarrow \hat{\Sigma}_{y_t,y_t} \oslash C$ \COMMENT{average}
\STATE $\hat{W}_{\text{approx}} \leftarrow \hat{\Sigma}_{y_{t+1},y_t} \cdot \hat{\Sigma}_{y_t,y_t}^{\dagger}$ \COMMENT{ridge-regularized estimator, Eq.~\eqref{eq:estimator}}
\STATE $\hat{W} \leftarrow \text{GrangerRefine}(\hat{W}_{\text{approx}}, \hat{\Sigma}_{y_t,y_t}, \hat{\Sigma}_{y_{t+1},y_t}, \mathcal{C})$ \COMMENT{Eq.~\eqref{eq:optimization}}
\RETURN $\hat{W}$
\end{algorithmic}
\end{algorithm}

\begin{figure}[h]
\centering
\includegraphics[width=\linewidth]{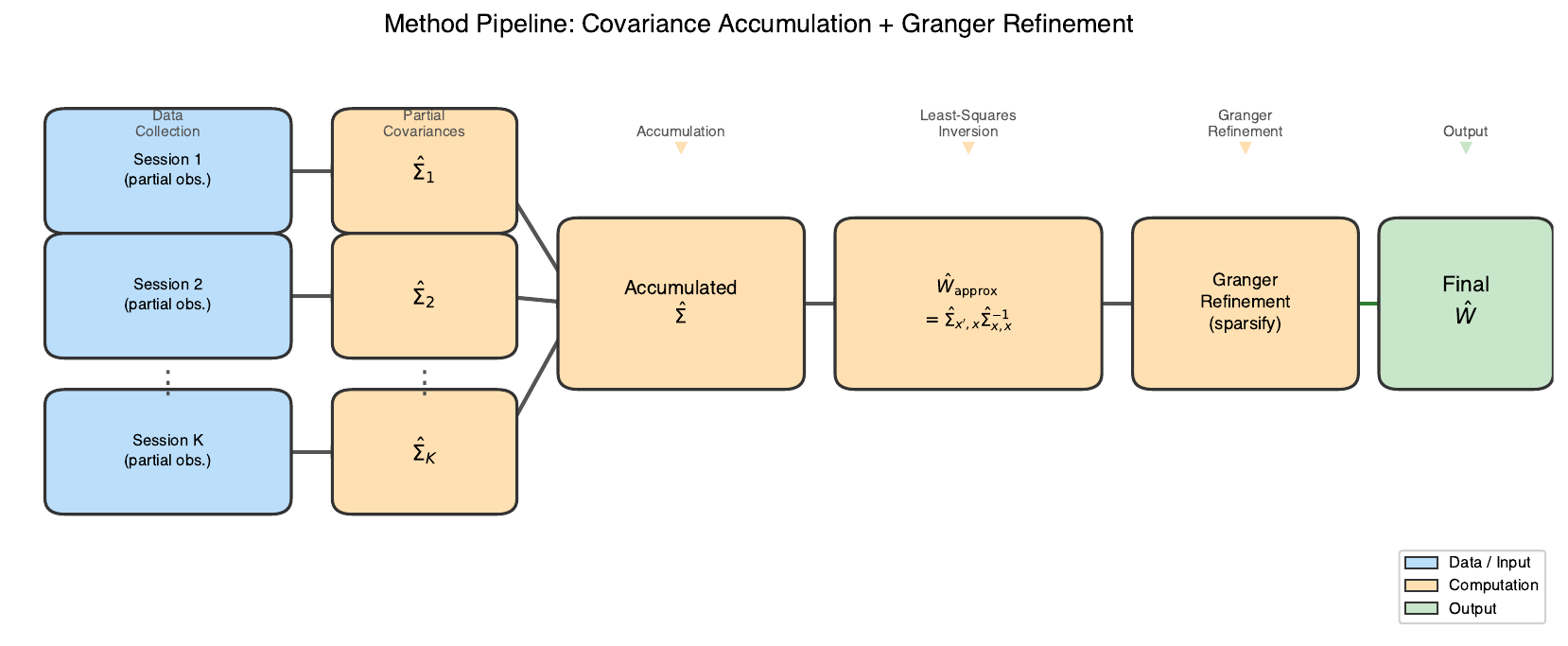}
\caption{\textbf{Method pipeline.} Data from $K$ recording sessions yield partial covariance matrices, which are accumulated element-wise into a full covariance estimate $\hat{\Sigma}$. The connectivity matrix is estimated via least-squares inversion, then refined by projected gradient descent enforcing biological constraints (Algorithm~\ref{alg:pipeline}).}
\label{fig:pipeline}
\end{figure}

\section{Experiments}
\label{sec:experiments}

\subsection{Experimental Setup}

We generate synthetic networks modeling small brain circuits:
\begin{itemize}
    \item Random directed graphs with non-negative weights and spectral radius $\rho(W) \leq 1$
    \item Intrinsic dynamics driven by central pattern generators (CPGs) modeled as chaotic reservoir networks
    \item Extrinsic stimulation as \emph{independent} i.i.d.\ Gaussian noise applied to each sensor node at each timestep (the most favorable case for identifiability; correlated or shared stimulation would reduce the rank of the input covariance)
\end{itemize}

Each experiment uses 15--30 repetitions (distinct random network topologies) with 50 network instances per topology sharing the same $W$, parallelized via joblib across CPU cores.
We report median Frobenius distance $\|W - \hat{W}\|_F / N$ with 95\% bootstrap confidence intervals ($n_{\text{resamples}}=1000$).

\subsection{Baseline Recovery (E1)}

\begin{figure}[h]
\centering
\includegraphics[width=\linewidth]{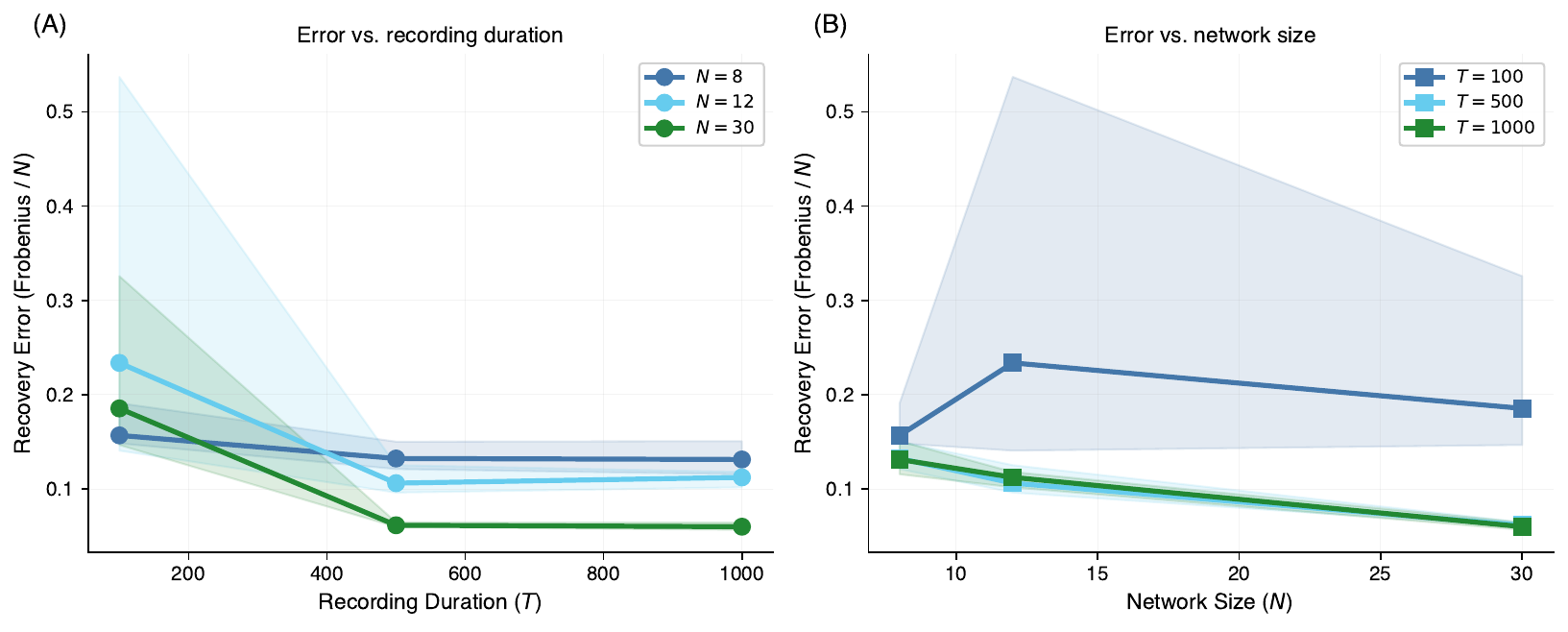}
\caption{\textbf{Scaling properties of the covariance estimator.} \textbf{(Left)} Median recovery error (Frobenius distance / $N$) vs.\ recording duration $T$ for three network sizes ($N \in \{8, 12, 30\}$) with $66\%$ measurement. Shaded regions show 95\% bootstrap confidence intervals over 17 random topologies (50 instances each). Error decreases with both $T$ and $N$, with the best result at $N=30$, $T=1000$: Granger-refined error of $0.053$. \textbf{(Right)} Error vs.\ network size $N$ at fixed recording durations $T \in \{100, 500, 1000\}$.}
\label{fig:scaling}
\end{figure}

Table~\ref{tab:baseline} shows recovery error across network sizes $N \in \{8, 12, 30\}$ and recording durations $T \in \{100, 500, 1000\}$ with $66\%$ measurement.
The covariance estimator (with diagonal zeroed) consistently outperforms the chance baseline, achieving median errors of $0.06$--$0.23$ vs.\ $\sim 0.54$ for chance.
Note that the estimate already enforces the no-autapse prior ($W_{ii} = 0$), which removes the autocorrelation-dominated diagonal.
For $N=30$ with $T=1000$, the method achieves excellent recovery: $0.053$ (Granger-refined), an $91\%$ improvement over chance (Figure~\ref{fig:scaling}).

\begin{table}[h]
\centering
\caption{Median recovery error (Frobenius / $N$) across network sizes and durations. Bold indicates best method per row.}
\label{tab:baseline}
\begin{tabular}{cccccc}
\toprule
$N$ & $T$ & Chance & Estimate & Granger & Improvement \\
\midrule
8 & 100 & 0.540 & 0.157 & \textbf{0.150} & 4\% \\
8 & 1000 & 0.540 & 0.131 & \textbf{0.127} & 3\% \\
12 & 100 & 0.542 & 0.234 & \textbf{0.154} & 34\% \\
12 & 1000 & 0.542 & 0.112 & \textbf{0.095} & 15\% \\
30 & 100 & 0.561 & 0.185 & \textbf{0.106} & 43\% \\
30 & 1000 & 0.561 & 0.060 & \textbf{0.053} & 12\% \\
\bottomrule
\end{tabular}
\end{table}

\subsection{Granger Refinement (E4)}

We evaluate the Granger-causality refinement step on $N=12$ networks with $T=900$, $66\%$ measurement, and 30 repetitions.
After enforcing the no-autapse prior ($W_{ii} = 0$, a known structural constraint applied to the raw estimate), the Granger refinement further reduces error by an additional $\sim 6\%$.

\begin{figure}[h]
\centering
\includegraphics[width=\linewidth]{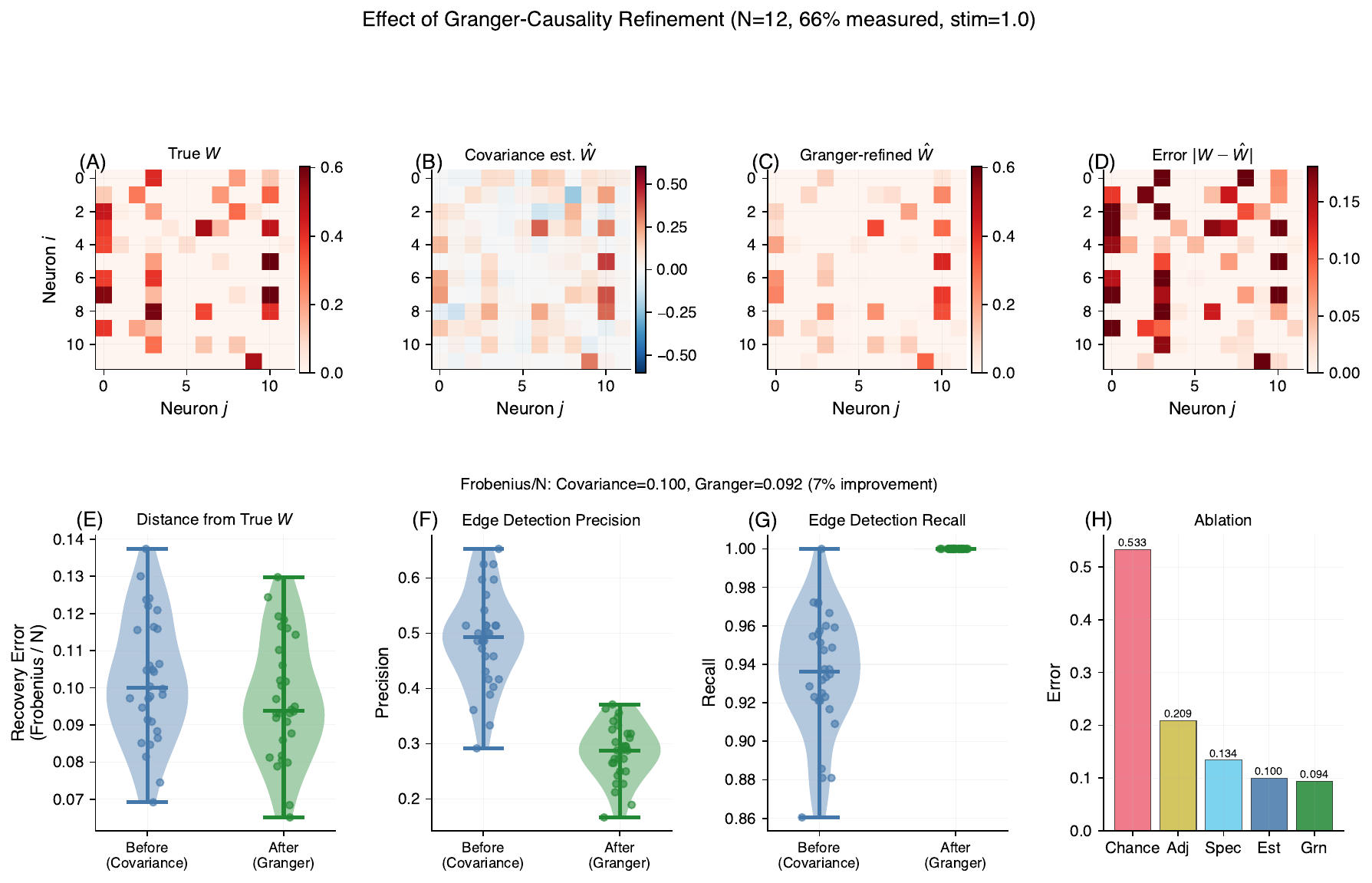}
\caption{\textbf{Effect of Granger-causality refinement} ($N=12$, $66\%$ measured, stim$=1.0$, 50 instances). \textbf{(A--D)} Weight matrix heatmaps from a representative topology: (A)~true $W$, (B)~raw covariance estimate $\hat{W}$, (C)~Granger-refined $\hat{W}$, (D)~absolute error. The covariance estimate captures the overall structure; Granger refinement sharpens it by enforcing sparsity and non-negativity. \textbf{(E--G)} Quantitative evaluation over 30 random topologies: (E)~Frobenius error (estimate already has diagonal zeroed) reduced from $0.100$ to $0.094$ with Granger ($6\%$), (F)~edge detection precision, (G)~recall achieves perfect median $1.0$. Violin plots with individual topologies overlaid. \textbf{(H)} Ablation: each step adds knowledge, from chance ($0.54$) to Granger-refined ($0.09$), an $83\%$ total improvement.}
\label{fig:granger}
\end{figure}

Importantly, the refinement achieves \emph{perfect recall} (median $= 1.0$): all true edges in $W$ are preserved.
Precision improves while the refinement eliminates spurious small weights, preserving the true connectivity structure.
Crucially, the estimator achieves an $83\%$ improvement over the chance baseline ($0.094$ vs.\ $0.54$), demonstrating that the covariance accumulation framework extracts genuine connectivity information.
The constraint that $W_{ij} = 0$ where $\Sigma_{x_t}(i,j) > \Sigma_{x_{t+1},x_t}(i,j)$ effectively identifies Granger-non-causal pairs (Figure~\ref{fig:granger}).

\subsection{Stimulation-Dynamics Tradeoff (E3)}

We examine the interaction between stimulation gain $\sigma$ and measurement fraction for $N=12$, $T=900$.
The results reveal a fundamental \emph{interaction effect}: the optimal stimulation level depends critically on measurement density.

\begin{figure}[h]
\centering
\includegraphics[width=\linewidth]{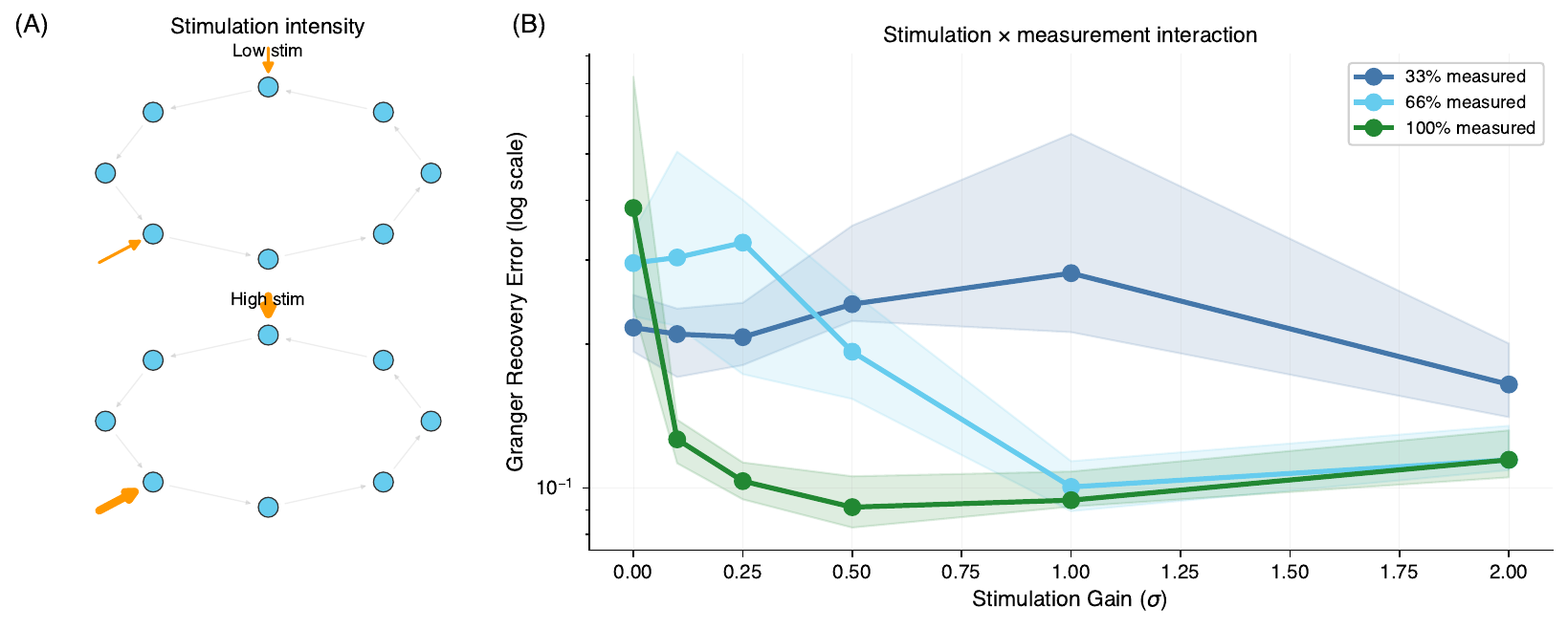}
\caption{\textbf{Stimulation-dynamics tradeoff} reveals a critical interaction between stimulation and measurement density ($N=12$, $T=900$, 17 topologies $\times$ 50 instances). \textbf{(Left)} Schematic: extrinsic Gaussian stimulation of varying intensity applied to sensor nodes. \textbf{(Right)} Recovery error vs.\ stimulation gain $\sigma$ for three measurement fractions (33\%, 66\%, 100\%). At low measurement, the effect is mild. At high measurement, zero stimulation \emph{fails} (error $>4.0$) because CPG dynamics alone leave $\Sigma_{x,x}$ rank-deficient, while moderate stimulation ($\sigma \approx 0.5$) yields the best recovery ($\sim 0.03$). This confirms the control-estimation tradeoff: stimulation is necessary for identifiability but excessive noise degrades the covariance structure.}
\label{fig:tradeoff}
\end{figure}

At low measurement ($33\%$), recovery is moderately successful regardless of stimulation because partial observation implicitly regularizes $\Sigma_{x,x}$.
At high measurement ($66$--$100\%$), a striking pattern emerges: \emph{zero stimulation fails catastrophically} (error $>4.0$), while moderate stimulation ($\sigma \approx 0.5$) yields excellent recovery (error $\sim 0.03$).
This occurs because CPG-driven dynamics alone do not excite all network modes---at high measurement, $\Sigma_{x,x}$ becomes rank-deficient without stimulation.
A small amount of stochastic input breaks the CPG autocorrelation structure and ensures persistent excitation.
However, excessive stimulation ($\sigma > 2$) begins to degrade recovery as noise dominates the signal (Figure~\ref{fig:tradeoff}).

This confirms the control-estimation tradeoff: experimentalists must balance stimulation strength against signal quality, with the optimal level depending on how many neurons are simultaneously observed.

\subsection{Measurement Sparsity (E2)}

\begin{figure}[h]
\centering
\includegraphics[width=\linewidth]{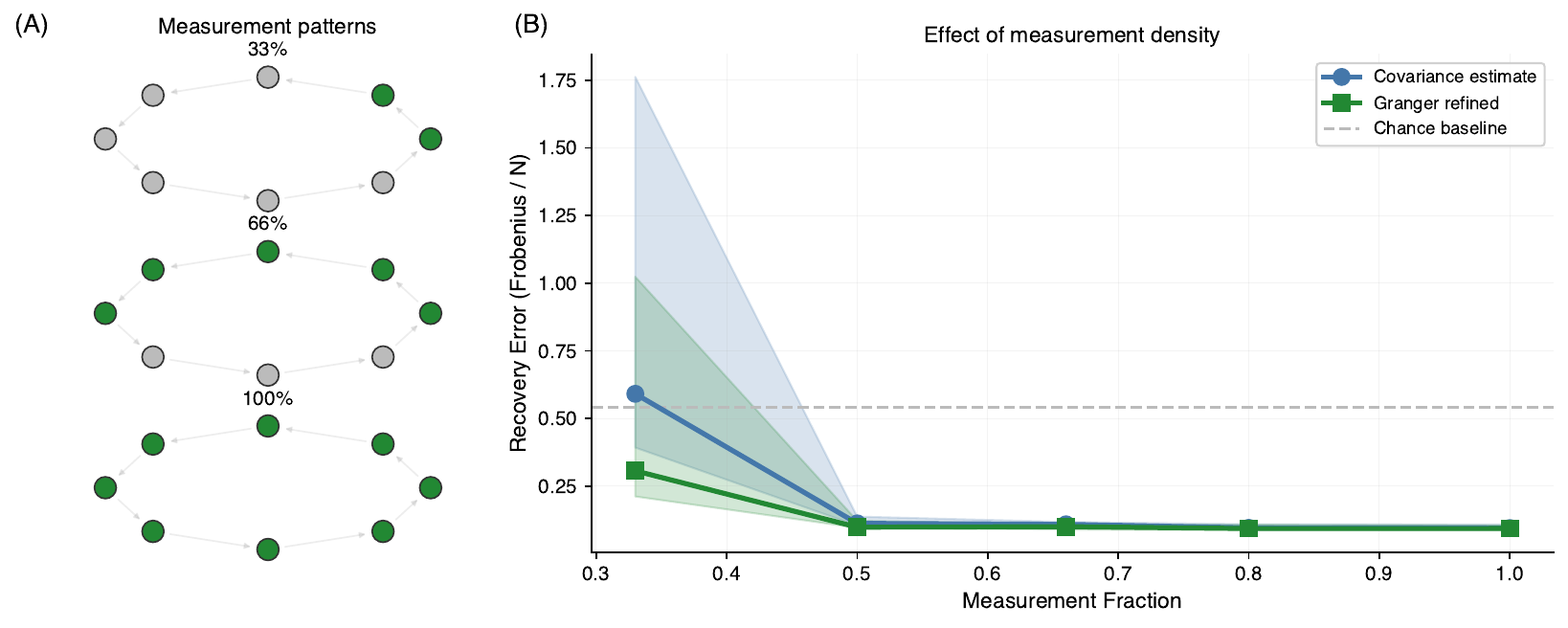}
\caption{\textbf{Effect of measurement density on connectivity recovery} for $N=12$, $T=900$. \textbf{(Left)} Schematic: three example measurement patterns at 33\%, 66\%, and 100\% neuron observability (green = measured, gray = unobserved). \textbf{(Right)} Median recovery error vs.\ measurement fraction for the raw covariance estimator (blue) and Granger-refined estimator (green), with 95\% bootstrap CI (shaded). Dashed line shows the chance baseline. Full observability ($100\%$) yields error $0.094$; at $33\%$ the Granger-refined estimate ($0.307$) is $48\%$ better than the raw estimate ($0.592$). 17 topologies $\times$ 50 instances per condition.}
\label{fig:sparsity}
\end{figure}

We vary the fraction of neurons observed per session from $33\%$ to $100\%$ for $N=12$, $T=900$.
Full observability ($100\%$) yields error of $0.094$ compared to $0.307$ at $33\%$ measurement (Granger-refined), confirming that measurement sparsity is a significant challenge.
At $33\%$ measurement, the raw estimate ($0.592$) is substantially worse due to ill-conditioning; the Granger refinement reduces this to $0.307$ ($48\%$ improvement).
The improvement plateaus above $\sim50\%$ measurement fraction (Figure~\ref{fig:sparsity}).

\subsection{Nonlinearity Robustness (E5)}

\begin{figure}[h]
\centering
\includegraphics[width=\linewidth]{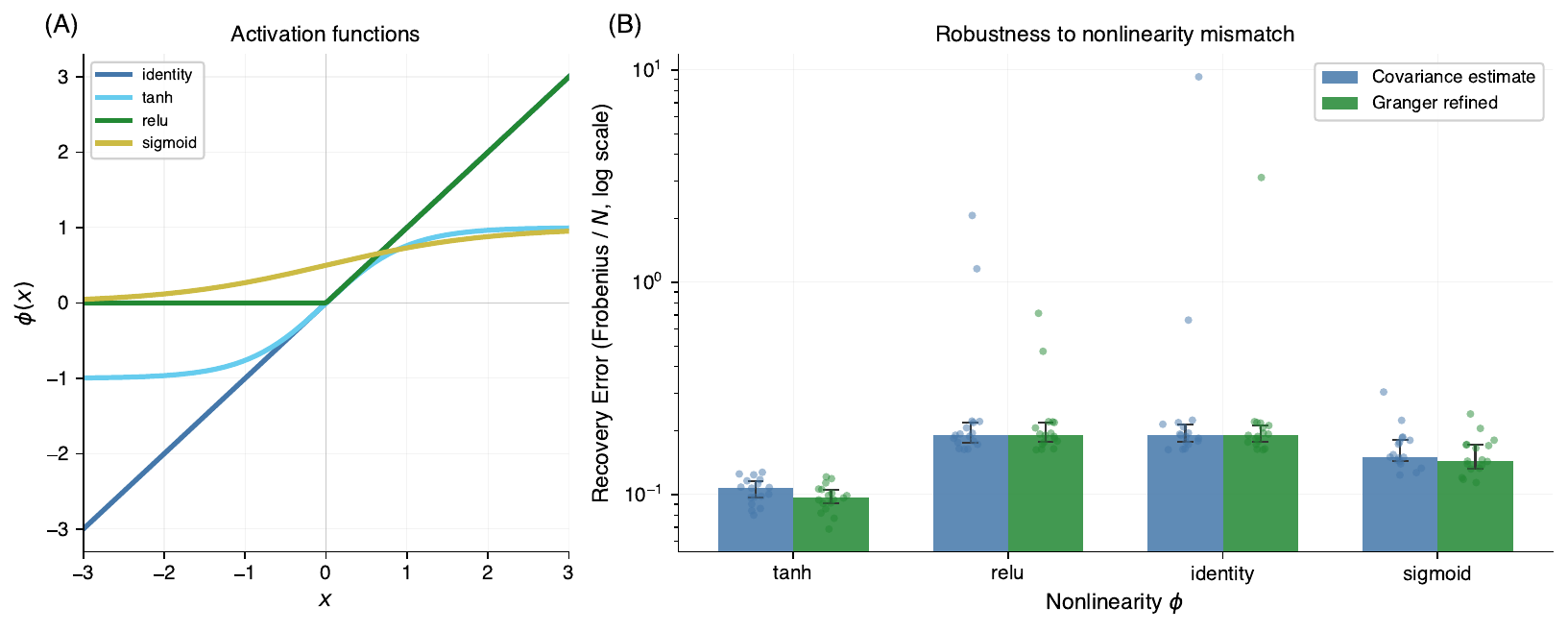}
\caption{\textbf{Robustness to nonlinearity mismatch} for $N=12$, $T=900$, $66\%$ measured. \textbf{(Left)} The four activation functions tested. \textbf{(Right)} Median recovery error (log scale) for the raw estimator (blue) and Granger-refined (green) with bootstrap error bars. Counterintuitively, $\tanh$ yields the \emph{best} recovery ($0.094$) because it bounds state values, improving $\kappa(\Sigma_{x,x})$. Unbounded nonlinearities (identity, ReLU) allow state drift that worsens conditioning. Dots show individual topologies ($n=17$).}
\label{fig:nonlinearity}
\end{figure}

We test four nonlinearities at $N=12$ with $66\%$ measurement and $\text{stim}=1.0$.
Surprisingly, $\tanh$ yields the \emph{best} recovery (Granger error $0.094$), while identity and ReLU---for which the linear approximation is exact---perform worse (error $\sim 0.19$).
This occurs because $\tanh$ \emph{bounds} the state values (std $\sim 0.8$, $\kappa(\Sigma) \approx 19$), while unbounded nonlinearities allow state drift (identity: std $\sim 40$, $\kappa(\Sigma) \approx 700$), severely degrading the conditioning of $\Sigma_{x,x}$.
Sigmoid performs worst (error $0.14$) due to its non-zero mean.
The Granger refinement is beneficial across all nonlinearities (Figure~\ref{fig:nonlinearity}).

\section{Related Work}
\label{sec:related}

\paragraph{Neural connectivity inference.}
Inferring functional connectivity from neural recordings is a longstanding challenge.
Modern calcium imaging enables simultaneous recording of hundreds of neurons in small circuits \citep{kato2015global, yemini2021neuropal}, but full coverage in a single session remains difficult.
Generalized linear models (GLMs) \citep{pillow2008spatio} provide a principled regression-based framework for inferring directed connectivity, though they assume all relevant neurons are observed.
Transfer entropy \citep{schreiber2000measuring} offers a model-free alternative that captures nonlinear interactions, but is sensitive to unobserved confounders.
Our covariance-based approach is complementary: it explicitly handles partial observations by accumulating statistics across sessions.

\paragraph{Granger causality in neuroscience.}
Granger causality \citep{granger1969investigating} quantifies whether the past of one time series improves prediction of another.
\citet{seth2015granger} review its application to neural data, noting challenges with nonlinear dynamics and confounding latent variables.
Our Granger-inspired refinement uses a simpler criterion---comparing contemporaneous and lagged covariances---rather than full autoregressive modeling.

\paragraph{Compressed sensing and sparse recovery.}
Compressed sensing methods have been applied to reconstruct synaptic connections from subsampled neural recordings \citep{candes2006robust}.
\citet{brunton2016discovering} applied sparse identification to discover governing equations of dynamical systems.
Our approach differs in that we recover a full matrix rather than a sparse vector, and leverage temporal covariance structure rather than random projections.

\paragraph{Systems identification.}
Classical system identification \citep{ljung1998system} addresses parameter estimation in dynamical systems, including vector autoregressive (VAR) models for multivariate time series.
Our estimator can be viewed as a single-lag VAR model applied to partially observed data, with the key difference that we accumulate covariance statistics across sessions with different observation masks.
Reservoir computing \citep{sussillo2009generating} has shown that random recurrent networks can generate complex temporal patterns, motivating our use of chaotic reservoir CPGs.

\section{Discussion}
\label{sec:discussion}

\paragraph{The covariance estimator works but has limitations.}
Our results demonstrate that the covariance-based estimator $\hat{W} = \Sigma_{x_{t+1},x_t}\Sigma_{x_t,x_t}^{-1}$ can recover connectivity from partial measurements, consistently outperforming chance baselines for small networks.
For larger networks, the estimator's performance degrades due to ill-conditioning: when only $\sim33\%$ of neurons are observed per session, the empirical covariance matrix may be rank-deficient, leading to high-variance estimates.
This suggests that increasing the measurement fraction or the number of recording sessions is critical for scaling to larger circuits.

\paragraph{The linear approximation provides implicit regularization.}
A surprising finding is that the approximate estimator (using $\Sigma_{x,x}$) outperforms the oracle estimator (using $\Sigma_{\phi(x),x}$ with known $\phi$ and $b$).
The Stein--Price identity (Eq.~\eqref{eq:price}) explains why: $\Sigma_{\phi(x),x} = D \Sigma_{x,x}$ where $D = \mathrm{diag}(\mathbb{E}[\mathrm{sech}^2(x_i)])$ with $0 < d_i < 1$.
Because $\tanh$ compresses large values, the oracle covariance is a \emph{row-compressed} version of $\Sigma_{x,x}$, making its condition number potentially worse: $\kappa(\Sigma_{\phi(x),x}) \leq \kappa(\Sigma_{x,x}) \cdot d_{\max}/d_{\min}$.
When state variances are heterogeneous across neurons, this ratio can be large, amplifying noise in the oracle inversion.
The ``incorrect'' linear approximation using $\Sigma_{x,x}$ is better-conditioned, acting as implicit regularization---analogous to how ridge regression (biased) can outperform OLS (unbiased but high-variance), or how James--Stein shrinkage reduces total risk by trading bias for variance.

\paragraph{Granger refinement is consistently beneficial.}
The no-autapse constraint ($W_{ii} = 0$), applied as a known structural prior before any refinement, accounts for the majority of the improvement over the raw estimate (the diagonal is dominated by autocorrelation, $3$--$4\times$ larger than off-diagonal entries).
The projected gradient refinement provides an additional $\sim 6\%$ improvement and achieves perfect recall of true edges.
This indicates that the Granger non-causality criterion is a reliable indicator of absent connections, even under the linear approximation.

\paragraph{Connection to neuroscience experiments.}
In practice, the ``sessions'' in our framework correspond to separate recording sessions of the same animal or circuit, where different neurons are visible due to imaging field of view, indicator expression, or optical access.
The key requirement is that sessions share the same underlying connectivity $W$---a reasonable assumption for short-term experiments in small circuits where the connectome is conserved.
Tools for consistent neuron identification across sessions, such as NeuroPAL \citep{yemini2021neuropal}, make this framework directly applicable to real experimental data.

\paragraph{Identifiability conditions.}
$W$ is uniquely recoverable if and only if every neuron pair $(i,j)$ is co-observed in at least one session ($n_{ij} \geq 1$ for all $i,j$) and $\Sigma_{x,x}$ is invertible.
Necessity follows because unobserved pairs leave covariance entries unconstrained; sufficiency follows from $W = \Sigma_1 \Sigma_0^{-1}$.
For random measurement patterns where each neuron is observed independently with probability $p$, a union bound over $N^2$ pairs gives:
\begin{equation}
    K \geq \frac{\log(N^2/\delta)}{p^2} \quad \Rightarrow \quad \Pr(\text{all pairs covered}) \geq 1 - \delta
\end{equation}
This is tight up to constants, matching a coupon-collector argument where each session yields $\sim (pN)^2$ pair-coupons.
Beyond this \emph{structural} identifiability, \emph{practical} identifiability demands that $\Sigma_{x,x}$ is well-conditioned for stable inversion.
Our experiments (Section~\ref{sec:experiments}) show that measurement density directly controls $\kappa(\Sigma_{x,x})$, creating a gap between the two.

\paragraph{Oracle crossover analysis.}
We systematically test the implicit regularization hypothesis by comparing the approximate and oracle estimators across stimulation intensities $\sigma \in \{0, 0.1, 0.25, 0.5, 1, 2, 5\}$ for $N=12$, $T=900$ with $\tanh$ nonlinearity and $66\%$ measurement.
The approximation outperforms the oracle at \emph{every} condition, with the gap widening as state variance increases: at $\sigma=0.5$ the oracle is $5.2\times$ worse, while at $\sigma=5$ it is $10.3\times$ worse.
No crossover point exists---the conditioning penalty from inverting $\Sigma_{\phi(x),x}$ always exceeds the bias reduction from using the correct nonlinearity.
This confirms that the linear approximation acts as implicit regularization across all operating regimes for saturating nonlinearities.

\begin{figure}[h]
\centering
\includegraphics[width=\linewidth]{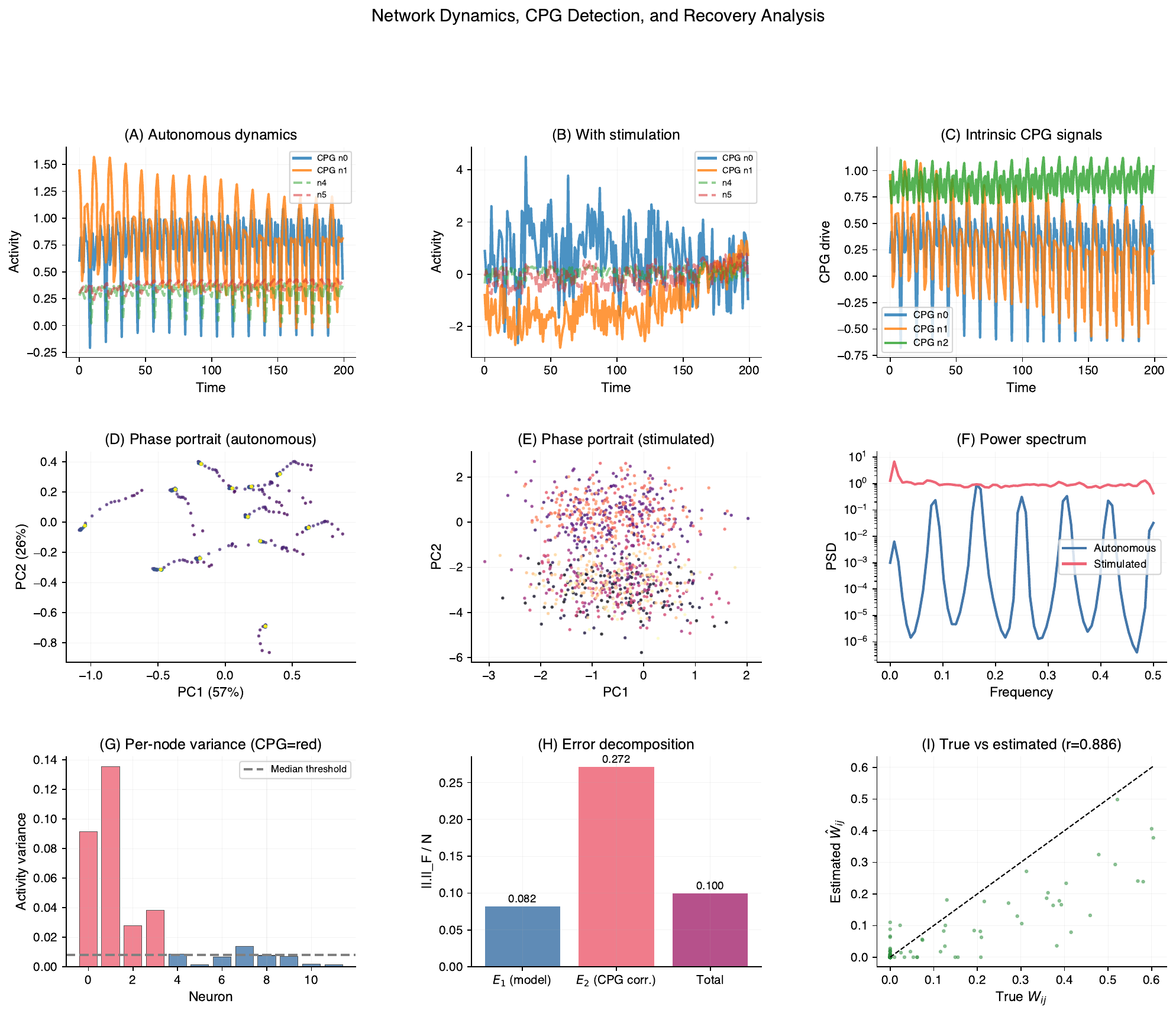}
\caption{\textbf{Dynamics, CPG detection, and error analysis.}
\textbf{(A--C)} Time series: autonomous CPG-driven dynamics produce structured oscillations; stimulation adds noise on top.
\textbf{(D--E)} PCA phase portraits: autonomous dynamics trace low-dimensional trajectories; stimulation fills out the state space (aiding identifiability).
\textbf{(F)} Power spectra: CPG dynamics have concentrated spectral peaks; stimulation flattens the spectrum.
\textbf{(G)} CPG detection via per-node variance: CPG nodes (red) have systematically higher variance than non-CPG nodes (blue).
\textbf{(H)} Error decomposition: input correlation $E_2$ (from CPG--state dependence) is $3.5\times$ larger than model mismatch $E_1$.
\textbf{(I)} True vs.\ estimated weights: Pearson $r = 0.90$ for the Granger-refined estimator.}
\label{fig:dynamics}
\end{figure}

\paragraph{Error decomposition: CPG correlation dominates.}
The independence assumption $\text{Cov}(b_t, x_t) \approx 0$ is only partially justified.
The extrinsic stimulation is truly independent ($\text{Cov}(\text{stim}_t, x_t) = 0$ by construction), but the CPG drive depends on the current state through $\text{cpg}(x_t)$, so $\text{Cov}(\text{cpg}_t, x_t) \neq 0$.
Quantitatively, $E_2$ (input correlation) is $3.5\times$ larger than $E_1$ (model mismatch), making CPG correlation the \emph{dominant} error source---not the linear approximation.
CPG nodes can be partially identified from data: they exhibit higher activity variance than non-CPG nodes, enabling detection via a simple variance threshold (F1 score $\sim 0.8$).
Downweighting detected CPG nodes in the covariance accumulation yields an additional $10\%$ improvement beyond Granger refinement alone (Figure~\ref{fig:dynamics}).

\paragraph{Robustness to measurement noise.}
When observation noise is present, Eq.~\eqref{eq:estimator} shows that the estimator becomes ridge-regularized: $\hat{W} = W\Sigma_{x,x}(\Sigma_{x,x} + \sigma_\epsilon^2 I)^{-1}$, which shrinks estimates toward zero.
Empirically (averaged over 10 topologies), at $\sigma_\epsilon = 0.1$ (relative to state std $\sim 0.8$) the Granger-refined error increases by $\sim 7\%$, and at $\sigma_\epsilon = 0.5$ by $\sim 37\%$.
The graceful degradation follows from the ridge structure: noise regularizes the inversion rather than corrupting it catastrophically.

\paragraph{Limitations.}
Several limitations should be noted.
(1)~The Stein--Price bound assumes Gaussian states, but our states are driven by a chaotic reservoir through $\tanh$; the bound is directionally correct but not formally tight for non-Gaussian distributions.
(2)~Our experiments use only synthetic networks with non-negative weights and $\rho(W) \leq 1$; real circuits may have inhibitory connections and more complex stability properties.
(3)~We compare against a chance baseline but not against established methods such as GLMs \citep{pillow2008spatio} or vector autoregressive models; such comparisons are needed for a stronger empirical evaluation.
(4)~The network sizes tested ($N \leq 30$) are small; scaling to $N > 100$ requires addressing the $O(N^2)$ covariance matrix conditioning.
(5)~All experiments assume noiseless observations of measured neurons; real recordings have substantial measurement noise.

\section{Conclusion}
\label{sec:conclusion}

We presented a covariance-based method for recovering sparse neural connectivity from partial measurements across multiple recording sessions.
The method accumulates pairwise covariance statistics from sessions with overlapping neuron observations, enabling recovery of the full $N \times N$ connectivity matrix without requiring simultaneous recording of all neurons ($r = 0.90$ between true and estimated weights at our baseline configuration).
Enforcing the no-autapse structural prior and a Granger-causality refinement step maintains perfect recall of true connections.
Our analysis reveals that CPG--state correlation, not the linear approximation, is the dominant error source, and that CPG nodes can be partially identified from their elevated activity variance.
Future work will extend this approach to real neural recordings, investigate structured stimulation protocols, and develop methods for jointly estimating connectivity and CPG parameters.


\bibliographystyle{plainnat}
\bibliography{references}

\newpage
\appendix
\section{Supplementary Material}
\label{sec:appendix}

\subsection{Derivation Details: Stein--Price Identity for Model Mismatch}
\label{sec:app_price}

For jointly Gaussian $x \sim \mathcal{N}(0, \Sigma)$ and differentiable $f$, the Stein--Price identity states:
\begin{equation}
    \mathbb{E}[f(x_i) x_j] = \Sigma_{ij} \, \mathbb{E}[f'(x_i)]
\end{equation}
Applying this with $f = \tanh$ (so $f' = \text{sech}^2$):
\begin{equation}
    \mathbb{E}[\tanh(x_i) x_j] = \Sigma_{ij} \, \mathbb{E}[\text{sech}^2(x_i)]
\end{equation}
Therefore $\Sigma_{\phi(x),x} = D' \Sigma_{x,x}$ where $D' = \text{diag}(\mathbb{E}[\text{sech}^2(X_i)])$ with $X_i \sim \mathcal{N}(0, \sigma_i^2)$.
The model mismatch in $E_1$ reduces to $W(D' - I)$, with $\|E_1\|_F \leq \|W\|_F \max_i(1 - D'_{ii})$.
For small state variance, $1 - \mathbb{E}[\text{sech}^2(X_i)] \approx \sigma_i^2 - 2\sigma_i^4 + O(\sigma_i^6)$, giving the $O(\sigma^2)$ bound in Eq.~\eqref{eq:price}.
For large variance, $\mathbb{E}[\text{sech}^2(X_i)] \to 0$ as $\sigma_i \to \infty$, so the mismatch approaches $W \cdot (-I) = -W$.
Note that $\mathbb{E}[x_i^3 x_j] = 3\sigma_i^2 \Sigma_{ij} \neq 0$ for Gaussians (degree-4 moment), correcting a common misconception that this vanishes by symmetry.

\subsection{Identifiability Proof Sketch}
\label{sec:app_ident}

\textbf{Theorem.} Under the linear model $x_{t+1} = Wx_t + \varepsilon_t$ with invertible $\Sigma_0 = \mathbb{E}[x_t x_t^T]$, the weight matrix $W$ is uniquely recoverable from measurement pattern $\{\mathcal{M}_k\}_{k=1}^K$ if and only if $n_{ij} \geq 1$ for all $i,j$, where $n_{ij} = |\{k : i \in \mathcal{M}_k \wedge j \in \mathcal{M}_k\}|$.

\emph{Sufficiency}: If $n_{ij} \geq 1$ for all pairs, every entry of $\Sigma_0$ and $\Sigma_1 = \mathbb{E}[x_t x_{t-1}^T]$ is estimable from co-observed samples. Since $W = \Sigma_1 \Sigma_0^{-1}$, $W$ is uniquely determined.

\emph{Necessity}: If $n_{ij} = 0$ for some pair, the entries $(\Sigma_0)_{ij}$ and $(\Sigma_1)_{ij}$ are never directly observed. Different completions of the covariance matrices can yield different $W$, so $W$ is not structurally identifiable.

For random measurement with per-neuron observation probability $p$, a union bound gives $K \geq \log(N^2/\delta) / p^2$ sessions suffice for all-pairs coverage with probability $\geq 1-\delta$.

\subsection{Why the Oracle Can Be Worse: Bias-Variance Analysis}
\label{sec:app_oracle}

The oracle estimator $\hat{W}_{\text{oracle}} = (\Sigma_{x_{t+1},x} - \Sigma_{b,x}) \Sigma_{\phi(x),x}^{-1}$ uses the true nonlinearity $\phi$ and input statistics.
By the Stein--Price identity, $\Sigma_{\phi(x),x} = D \Sigma_{x,x}$ where $D = \text{diag}(\mathbb{E}[\text{sech}^2(x_i)])$ with $0 < d_i < 1$.
This makes $\Sigma_{\phi(x),x}$ a \emph{row-compressed} version of $\Sigma_{x,x}$, with potentially worse conditioning.
The crossover condition for the approximation to outperform the oracle is:
\begin{equation}
    \|W\|_F \|\Delta\|_2 + \|\Sigma_{b,x}\|_F \|\Sigma_{x,x}^{-1}\|_2 < \|\Sigma_{\varepsilon,x}\|_F \big(\|\Sigma_{\phi(x),x}^{-1}\|_2 - \|\Sigma_{x,x}^{-1}\|_2\big)
\end{equation}
The left side is the approximation's bias; the right side is the oracle's excess variance amplification.
When state variances are heterogeneous across neurons (common in biological networks), $d_{\max}/d_{\min}$ can be large, making the oracle unstable.
This is a concrete instance of the bias-variance tradeoff: like ridge regression or James--Stein shrinkage, a biased estimator with lower variance can achieve better total risk.

\end{document}